\documentclass[a4paper]{jpconf}
\usepackage{graphicx}
\usepackage{wasysym}
\usepackage[numbers,square,sort&compress]{natbib}
\begin{document}
\title{Magnetic-field-induced 1st order transition to FFLO state at paramagnetic limit in 2D superconductors}

\author{N A Fortune$^1$, C C Agosta$^2$, S T Hannahs$^3$ and J A Schleuter$^4$}

\address{$^1$ Department of Physics, Smith College, Northampton MA, USA}
\address{$^2$ Department of Physics, Clark University, Worcester MA, USA}
\address{$^3$ National High Magnetic Field Laboratory, DC Field Facility, Tallahassee FL, USA}
\address{$^4$ Argonne National Laboratory, Materials Science Division, Argonne IL, USA}


\ead{nfortune@smith.edu}

\begin{abstract}
We have recently reported the first direct calorimetric observation   of a magnetic-field-induced first-order phase transition   into a high-field FFLO superconducting state at the Clogston-Chandrasekar `Pauli' paramagnetic limit$H_p$ in a 2D superconductor $\kappa-{\textrm{(BEDT-TTF)}}_2{\textrm{Cu}}{\textrm{(NCS)}}_2$. The high-field state is both higher entropy and strongly paramagnetic, as thermodynamically required for the FFLO state.  Here we  compare our results with theoretical predictions for the field dependence of the high-field FFLO state in the 2D limit, revealing tentative evidence for transitions between FFLO states of differing order parameter.  We also present calorimetric evidence for a 1st order phase transition into the FFLO state for a second 2D organic superconductor: ${\beta}^{\prime\prime}-{\textrm{(BEDT-TTF)}}_2{\textrm{SF}}_5{\textrm{(CH)}}_2{\textrm{(CF)}}_2{\textrm{(SO)}}_3$. 
\end{abstract}

\section{Introduction}
What is the highest possible applied magnetic field in which superconductivity can exist for an electronically 2D superconductor? One possible answer derives from the paramagnetic spin susceptibility of the conduction electrons. Because the electrons  in BCS superconducting pairs have oppositely aligned spins with momenta $(\mathbf{k\uparrow}, -\mathbf{k\downarrow})$, the reduction in Zeeman energy that arises from realignment of the electron spin in an applied magnetic field must eventually exceed the  reduction in electronic energy available from the formation of superconducting Cooper pairs as the magnetic field increases in strength. The critical magnetic field at which this should occur is known variously as the Clogston-Chadrasakar or Pauli paramagnetic limit $H_p$. 

In practice, orbital scattering from magnetic vortices typically limits superconductivity in  `clean' (impurity free) limit materials to magnetic fields much lower than $H_p$. For a 2D superconductor, however, this scattering can be suppressed by aligning the magnetic field exactly parallel to the 2D superconducting planes (thereby confining the vortices to the regions between the planes). This has been confirmed in angle-dependent studies of the Pauli-limited 2D organic superconductor $\alpha-{\textrm{(BEDT-TTF)}}_2{\textrm{NH}}_4{\textrm{(SCN)}}_4$ \cite{Coffey2010gl}. 

Perhaps surprisingly, then, it is possible for superconductivity to persist even above this expected paramagnetic limit. One way for this to happen is if Cooper pair formation above $H_p$ occurs  between Zeeman-split parts of the Fermi surface with momenta $(\mathbf{k\uparrow}, -\mathbf{k + q\downarrow})$, as first proposed by Fulde and Ferrell \cite{FULDE1964dq} and independently by Larkin and Ovchinnikov \cite{LARKIN1965uw}.

\begin{figure}[t]
\includegraphics[width=3in]{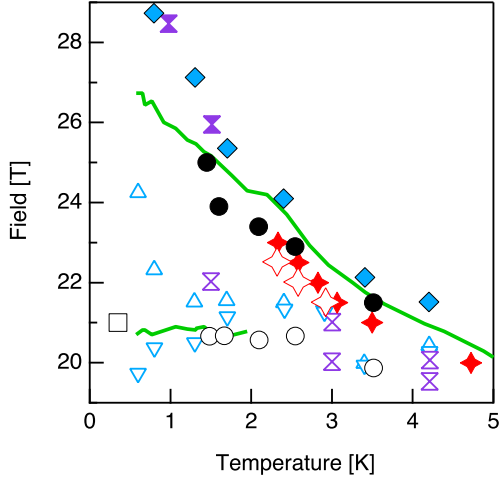}
\hspace{2pc}%
\begin{minipage}[b]{18pc}\caption{\label{fig:priorresults} Previously reported transition points for the $H_{\textrm{c2}}$ (solid symbols) and FFLO (hollow symbols) phase boundaries  for  $\kappa-{\textrm{(BEDT-TTF)}}_2{\textrm{Cu}}{\textrm{(NCS)}}_2$ for a magnetic field applied parallel to the 2D superconducting layers. Our penetration depth measurements \cite{Agosta2012wk} are shown as green lines. Calorimetric measurements by Lortz \cite{Lortz2007is} are shown as red stars. Magnetic-torque measurements by Bergk \cite{Bergk2011eh} and Tuschiya \cite{Tsuchiya2015bl} are shown as blue diamonds/ triangles and purple hourglass symbols, respectively. NMR measurements by Wright \cite{Wright2011ct} and Mayaffre \cite{Mayaffre2014dj} are shown as black squares and circles, respectively. } 
\end{minipage}
\end{figure}

This `FFLO' phase is only energetically favored to exist for  $H_p \leq H \leq H_{\textrm{c2}}$, where for a 2D superconductor, the zero temperature upper critical field ${{H}}_{c2}(0)$ has a predicted value of ${H}^{2D}_{c2}(0) \approx 1.42\ H_p$.
\cite{BULAEVSKII1973, Buzdin1997ct, Matsuda2007cv}. In addition, the temperature  cannot exceed a critical value $T^*$ where $T^* = 0.56\ T_c$ \cite{Buzdin1997ct, Shimahara1997bk}.   For a 2D superconductor like $\kappa-{\textrm{(ET)}}_2{\textrm{Cu}}{\textrm{(NCS)}}_2$ with a zero field superconducting critical temperature $T_c = 9.1 \pm 0.2{\textrm{\ K}}$ \cite{Lortz2007is, Agosta2012wk} and a Pauli limit $H_p = 20.7 \pm 0.4 \textrm{\ T}$ \cite{Wright2011ct, Agosta2012wk}, theory then predicts $T^* =5.1 \pm 0.1{\textrm{\ K}}$ and ${H}^{2D}_{c2}(0)= 29.5 \pm 0.7 \textrm{\ T} $. 

As noted by Matsuda and Shimahara \cite{Matsuda2007cv}, the $\mathbf{q}$ vector in the FFLO phase gives rise to spatial symmetry breaking, but because of the symmetry of the system, there is more than one equivalent $\mathbf{q}$ which gives the same upper critical field. In general, the superconducting order parameter $\Delta(\mathbf{r})$ at any particular magnetic field can be expressed as a linear combination of plane waves \cite{LARKIN1965uw}. 
The `FF' order parameter initially proposed by Fulde and Ferrell \cite{FULDE1964dq}  
corresponds to a single plane wave with $\Delta(\mathbf{r}) = \Delta_1 {\exp}^{i \mathbf{q \cdot r}}$, which results in  de-pairing of part of the Fermi surface at $H_p$  and an increase in entropy at the crossover from the BCS to the FF superconducting phase \cite{FULDE1964dq}. The `LO' order parameter initially proposed by Larkin and Ovchinnikov \cite{LARKIN1965uw} corresponds to the linear combination of two plane waves
with $\Delta(\mathbf{r}) = 2 \Delta_1 \cos(\mathbf{q \cdot r})$, resulting in a 1D spatial oscillation of the order parameter with wavelength $2 \pi / q$ and  planes of paramagnetic spins at the nodes. Here too there is an increase in entropy upon crossing over to the LO phase \cite{Cai2011gl, Tokiwa2012fg}. 

\section{Experimental results for $\kappa-{\textrm{(BEDT-TTF)}}_2{\textrm{Cu}}{\textrm{(NCS)}}_2$}

Previous calorimetric \cite{Lortz2007is} and magnetic torque \cite{Bergk2011eh, Tsuchiya2015bl} and  measurements on this material have presented thermodynamic evidence for a change in curvature of the superconducting to normal state phase boundary $H_{\textrm{c2}}(T)$ at what was inferred to be the high temperature onset of an FFLO superconducting phase. They have, in addition, reported direct observations of an FFLO phase transition, but as shown in Fig.~\ref{fig:priorresults}, the  locations, curvatures, and temperature dependences of the reported phase boundaries are in contradiction with each other, with penetration depth measurements \cite{Agosta2012wk}, and, at low temperature, with NMR measurements\cite{Wright2011ct, Mayaffre2014dj}. The hollow red stars hugging the $H_\textrm{c2}(T)$ phase boundary represent a 1st order transition initially thought to represent the crossing of the FFLO phase boundary \cite{Lortz2007is} but are now attributed \cite{Beyer2012be} to a small out of plane component of the magnetic field and the corresponding  emergence of a vortex phase near $H_\textrm{c2}$. The divergence  in the magnetic torque measurements may reflect the difficulty in determining strongly magnetic-field-angle-dependent phase boundaries with a measurement that requires the rotation of a sample in magnetic field.

\begin{figure}[t]
\includegraphics[width=3in]{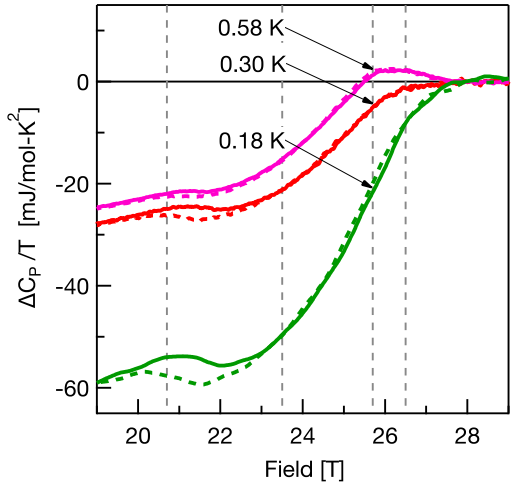}
\hspace{2pc}%
\begin{minipage}[b]{18pc}\caption{\label{fig:lowTdata} Magnetic-field induced change (solid for sweep up, dotted for sweep down) in the specific heat of $\kappa-{\textrm{(ET)}}_2{\textrm{Cu}}{\textrm{(NCS)}}_2$ for fields applied parallel to the superconducting layers. $\Delta C/T$ is set equal to zero above $H_{\textrm{c2}}$. A hysteretic first order phase transition at $H_p = 20.7 \pm 0.4 \textrm{\ T}$ into the FFLO superconducting phase followed by a rise in specific heat at the system approaches the normal state is visible at low temperature. The dashed vertical lines are guides to the eye marking $H_p$, the end of the hysteretic bubble in $C(H)$ above $H_p$, and breaks in the slope of $C(H)$ at fields corresponding to possible phase transitions within the FFLO region. 
 }
\end{minipage}
\end{figure}

\begin{figure}[t]
\begin{minipage}{18pc}
\includegraphics[width=3in]{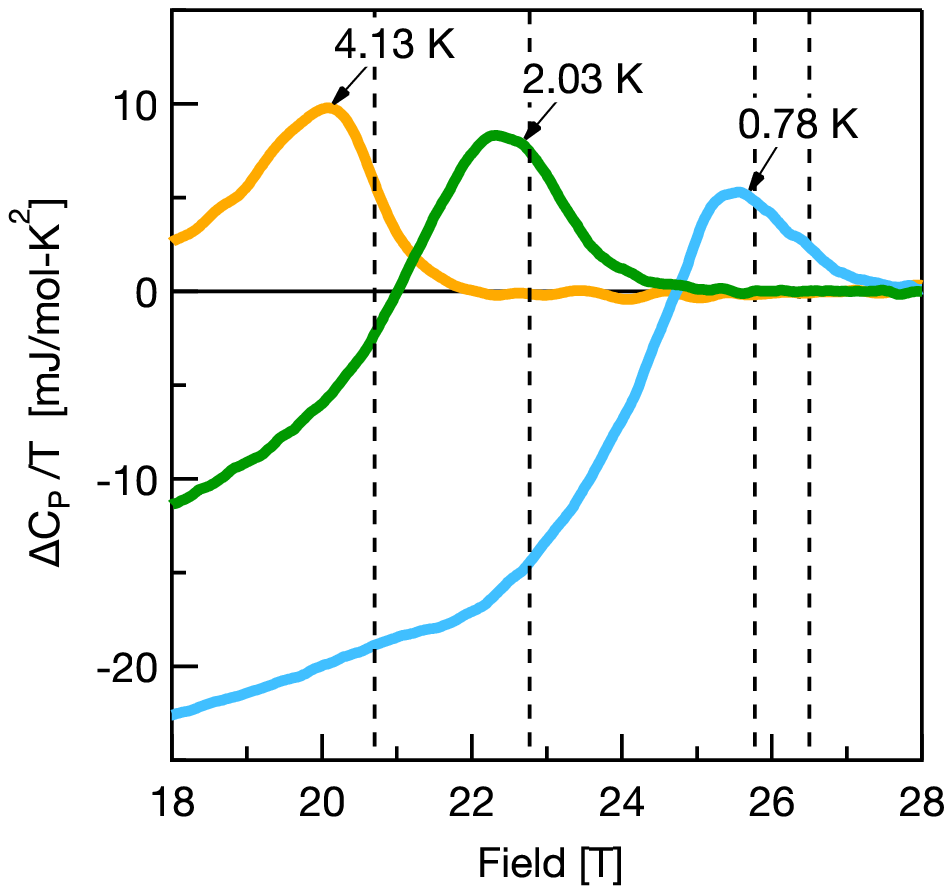}
\caption{\label{fig:highTdata}Relative change in the specific heat (scaled by temperature) of $\kappa-{\textrm{(BEDT-TTF)}}_2{\textrm{Cu}}{\textrm{(NCS)}}_2$ with magnetic field for fields applied parallel to the superconducting layers.  At higher temperatures, broad peak emerges 
at high field,
terminating at $H_\textrm{c2}(T)$. A strong  enhancement is seen in the NMR relaxation rate over the same field range at these temperatures \cite{Mayaffre2014dj}. The FFLO transition at $H_p$ is visible as a small bump at 0.78 K then as a slope change at higher T. Dashed lines are the same as in Fig.~\ref{fig:lowTdata}. Identified from changes in slope of $C(H)$ at lower T, they may correspond to the locations in field of  phase transitions 
to FFLO states of different symmetries, as predicted for FFLO superconductors in the 2D limit \cite{Shimahara1997bk}. 
}
\end{minipage}\hspace{2pc}
\begin{minipage}{18pc}
\includegraphics[width=3in]{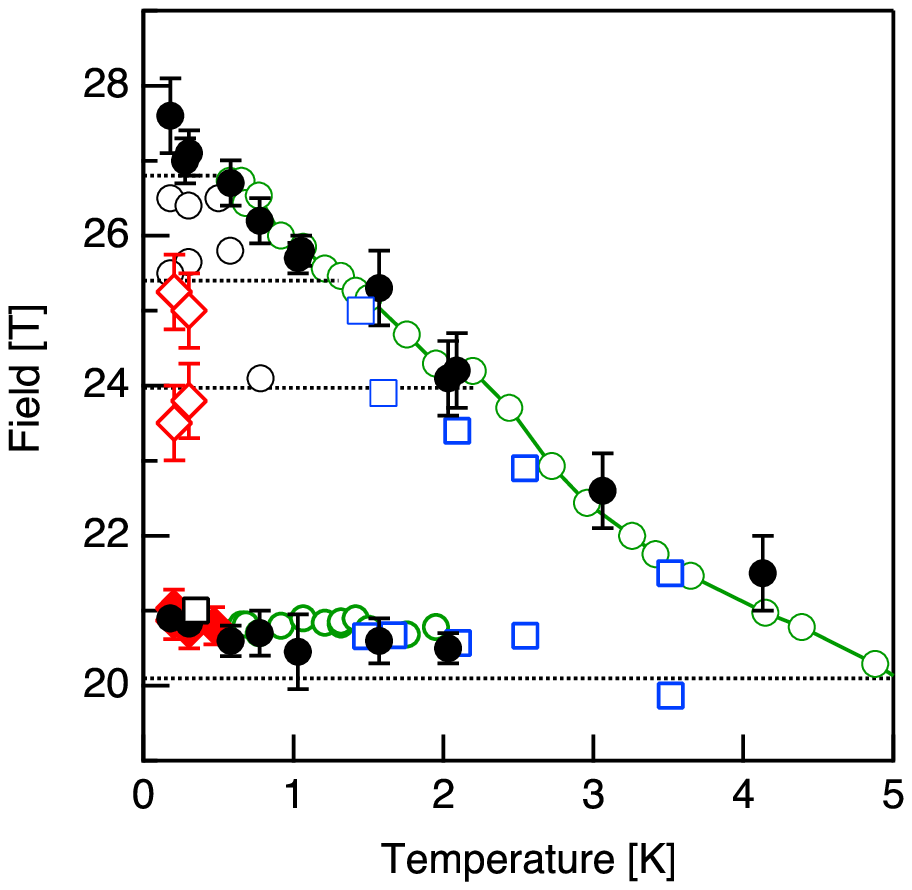}
\caption{\label{fig:goodresults} Phase diagram for $\kappa-{\textrm{(BEDT-TTF)}}_2{\textrm{Cu}}{\textrm{(NCS)}}_2$, showing locations of the  FFLO phase boundary at $H_p = 20.7\textrm{\ T}$ and $H_{\textrm{c2}}(T)$. Our calorimetric \cite{Agosta2017prl}, magnetocaloric \cite{Agosta2017prl}, and penetration depth \cite{Agosta2012wk} data are shown as solid black circles, solid red diamonds, and hollow green circles, respectively.  NMR results are shown as hollow black \cite{Wright2011ct} and blue \cite{Mayaffre2014dj} squares. Additional features in the calorimetric and magnetocaloric data are shown as hollow black circles and hollow red diamonds. Dashed lines represent predicted phase transitions \cite{Shimahara1997bk} for a 2D superconductor with 
s-wave pairing
to FFLO states of different symmetries. For d-wave pairing, only one transition within the FFLO state (at 26.5 T) should occur instead of the three shown here. 
}
\end{minipage} 
\end{figure}

For comparison, our own recent calorimetric measurements of the magnetic-field-dependent specific heat for $\kappa-{\textrm{(ET)}}_2{\textrm{Cu}}{\textrm{(NCS)}}_2$ for a magnetic field applied parallel to the 2D superconducting planes are shown in Figs.~\ref{fig:lowTdata} and \ref{fig:highTdata}. Like the NMR measurements, our measurements have the advantage of being able to use the angle dependence of $H_{c2}(\theta)$ to orient the sample in field \cite{Fortune2014cp}. At low temperatures, as shown in Fig.~\ref{fig:lowTdata}, a hysteretic first order phase transition at $H_p = 20.7 \pm 0.4 \textrm{\ T}$ into the FFLO superconducting phase followed by a rise in specific heat at the system approaches the normal state is visible, and as is shown the phase diagram presented in Fig.~\ref{fig:goodresults}, the calorimetrically determined locations of both transitions are in good agreement with NMR measurements \cite{Wright2011ct, Mayaffre2014dj}. Magnetocaloric measurements \cite{Agosta2017prl} indicate that the system above $H_p$ is paramagnetic, as expected for the FFLO state; the upward curvature of $C_p(H)$ is characteristic of strongly Pauli-paramagnetic superconductors at low T \cite{Machida2008cy}.   

Looking closer within the FFLO superconducting region, we find additional fine structure in the form of (1) a hysteretic bubble in the specific heat $ C_T(H)$ extending approximately 3 tesla above $H_p$ and (2) in the form of apparent changes in slope of $ C_T(H)$ just below the upper critical field phase boundary $H_{\textrm{c2}}$ and the field-independent normal state.   Interestingly, the collapse of the bubble in the 23 - 24 tesla region coincides with a pronounced drop in $|dM/dT|$ seen in the magnetocaloric effect. These phenomena are suggestive of an predicted change in the orientation of those spins at this field due to a change in the FFLO superconducting order parameter from 1D to 2D \cite{Shimahara1997bk}.  At higher temperatures, a  broad peak emerges at high field, terminating at $H_\textrm{c2}(T)$. A strong  temperature-dependent enhancement is seen in the NMR relaxation rate over the same field range at these temperatures \cite{Mayaffre2014dj}, indicating that the Schottky-like peaks may arise from strong temperature-dependent spin flip scattering within the superconducting state. 

\section{Predictions for field-induced FFLO states in a 2D superconductor}

The additional features seen in the specific heat and magnetocaloric effect above $H_p$ may indicate the existence of  phase transitions within the FFLO region. Several theoretical proposals  can be found in the literature \cite{Matsuda2007cv} but we restrict ourselves here to a comparison of our experimental results with that expected for a 2D superconducting film in the FFLO state \cite{Shimahara1997bk}. 

For a 2D superconducting film,  Shimahara \cite{Shimahara1997bk} predicts that  the originally proposed 1D planar spatially modulated `LO' order parameter \cite{LARKIN1965uw} $\Delta(\mathbf{r}) = 2 \Delta_1 \cos(\mathbf{q \cdot r})$ remains the lowest energy configuration just below $T^*$ 
(where $T_c$ is the zero field superconducting critical temperature and $T^*= 0.56\  T_c$ is the FFLO superconducting critical temperature )
but as the temperature lowers still further, a sequence of transitions to other FFLO phases with larger numbers of nodes are predicted to occur. The argument is that increasing the number of plane waves increases the number of nodes, thereby increasing the area over which the superconducting order parameter is small, which in turn leads to a reduction in the Pauli paramagnetic energy of the quasiparticles excited around the nodes  and a consequent gain in spin-polarization energy \cite{Shimahara1997bk, Matsuda2007cv}. 

For a superconductor with s-wave pairing, four different FFLO phases are expected, beginning with the 
1D planar LO phase for $ 0.24\ T_c \leq T \leq T^* = 0.56\  T_c$ followed by a 
state with triangular symmetry for $ 0.16\ T_c \leq T \leq 0.24\  T_c$, a 
state with square symmetry for $ 0.05\  T_c \leq T \leq 0.16\  T_c$, and finally a 
state with hexagonal symmetry for $  T \leq 0.05\  T_c$. For a superconductor with d-wave pairing,
 the more restrictive symmetry of the superconducting pairing means that only two phases are expected: the LO phase for $ 0.06\  T_c \leq T < T^* = 0.56\  T_c$ and the square state for $  T \leq 0.06\  T_c$. We can determine the fields 
 at which  field-induced transitions to the LO , triangle  $(\triangle)$, square  $(\Square)$, and/or  hexagonal  $(\hexagon)$ states should occur by finding the value $H_{\textrm{c2}}(T)$ corresponding to each of these temperatures. 
 
 As the nature of the superconducting pairing in $\kappa-{\textrm{(BEDT-TTF)}}_2{\textrm{Cu}}{\textrm{(NCS)}}_2$ is still a matter of debate \cite{cryst2020248}, we  consider each prediction in turn. For  $\kappa-{\textrm{(BEDT-TTF)}}_2{\textrm{Cu}}{\textrm{(NCS)}}_2$, Shimahara's theory for s-wave pairing predicts that field-induced transitions to various FFLO states should occur at  $H_{LO} 
 = 20.1\textrm{\ T}$, $H_{\triangle} 
 = 24.0\textrm{\ T}$, $H_{\Square} 
  = 25.4\textrm{\ T}$, and $H_{\hexagon} 
   = 26.8\textrm{\ T}$.  
   If we instead assume d-wave pairing, the transition to the 1D LO state again occurs at $H_{LO}  = 20.1\textrm{\ T}$ followed by a single transition with the FFLO state to a 2D square state at $H_{\Square} =  26.5\textrm{\ T}$.    
   
   The critical fields for the predicted transitions within the FFLO state for s-wave pairing are shown as dashed lines in Fig.~\ref{fig:goodresults}. As seen in Fig.~\ref{fig:goodresults}, there is at least a plausible correspondence between theory and experiment, suggesting that there is indeed a cascade of field-induced phase transitions to FFLO states with differing 1D and 2D order parameters (provided that the features identified here in the specific heat and the magnetocaloric effect correspond to  phase transitions and that the system can be properly modeled as a s-wave pairing superconductor in the 2D superconducting limit). Higher resolution measurements of specific heat and NMR as a function of field and temperature are needed to verify the tentative conclusion reached here. Of particular interest would be a comparison of measurements as a function of temperature for each of the proposed high field FFLO states.    

\section{Preliminary results for ${\beta}^{\prime\prime}-{\textrm{(BEDT-TTF)}}_2{\textrm{SF}}_5{\textrm{(CH)}}_2{\textrm{(CF)}}_2{\textrm{(SO)}}_3$}

\begin{figure}[b]
\includegraphics[width=3in]{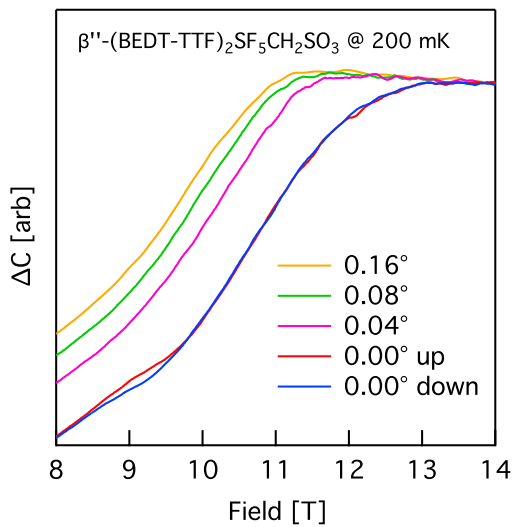}
\hspace{2pc}%
\begin{minipage}[b]{18pc}\caption{\label{fig:beta_angle_dependence} Dependence of the magnetic-field-dependent heat capacity of ${\beta}^{\prime\prime}-{\textrm{(BEDT-TTF)}}_2{\textrm{SF}}_5{\textrm{(CH)}}_2{\textrm{(CF)}}_2{\textrm{(SO)}}_3$ on magnetic field orientation at 200 mK. A first order transition to a field-induced FFLO superconducting state is observed for the $0^{\circ}$ plane parallel  orientation at a field consistent with a recent NMR determination for $H_{FFLO}$ of $9.3 \pm 0.1\textrm{\ T}$ \cite{Koutroulakis2016jpa}. The calorimetric signature of the hysteretic 1st order FFLO phase transition disappears for a change in sample orientation  $\Delta \theta$ as small as ${0.04}^{\circ}$. This change is also accompanied by a steep drop in the critical field, as expected for a 2D superconductor.}
\end{minipage}
\end{figure}

The calorimetric signature of the FFLO phase transition in $\kappa-{\textrm{(BEDT-TTF)}}_2{\textrm{Cu}}{\textrm{(NCS)}}_2$ is strongly field-orientation dependent, disappearing for a rotation of less than ${1}^{\circ}$ \cite{Agosta2017prl}, as expected for a 2D superconductor where the exclusion of magnetic vortices from the superconducting layers needed to achieve the FFLO state relies on a parallel orientation of the applied field relative to the layers.  If our interpretation is correct, however, then we should expect to see (1) corresponding first-order phase transitions at $H_p$ into FFLO states in related 2D organic superconductors in the clean impurity-scattering-free limit and (2) a strong field-angle-dependence of the FFLO states.  

As a test, we have carried out  preliminary results for angle-dependence of the heat capacity as a function of field $C(H)$ for the highly 2D superconductor  ${\beta}^{\prime\prime}-{\textrm{(BEDT-TTF)}}_2{\textrm{SF}}_5{\textrm{(CH)}}_2{\textrm{(CF)}}_2{\textrm{(SO)}}_3$ with $T_c = 4.3 \textrm{\ K}$ \cite{Beyer2012be}) at 200 mK. As seen in  Fig.~\ref{fig:beta_angle_dependence},  we observe for the field-parallel sample orientation a hysteric first order phase transition in a narrow region centered around $9.1 \pm 0.5 \textrm{\ tesla}$ followed by transition to the normal state at approximately 13 tesla, in agreement with expectation. We also find that  the calorimetric signature of the hysteretic 1st order FFLO phase transition disappears for a change in sample orientation  $\Delta$ as small as ${0.04}^{\circ}$ and that this change is also accompanied by a steep drop in the superconducting critical field. Although initial empirical determinations put $H_p$ and $H_{FFLO}$ for this material in the range of $9.7 \textrm{\ T}$ to  $10.5 \textrm{\ T}$ \cite{Cho2009jw, Agosta2012wk, Beyer2012be}, our lower value for $H_{FFLO}$ is consistent with both a recent NMR observation \cite{Koutroulakis2016jpa} of a phase transition from a uniform superconducting to FFLO superconducting state at $9.3 \pm 0.1\textrm{\ T}$ and the theoretical expectation of an FFLO phase transition at a field $H_{FFLO} \leq H_{\textrm{c2}}(0.56\ T_c) = 9.2 \pm 0.2 \textrm{\ T}$. Here $H_{\textrm{c2}}(0.56\ T_c)$  has been empirically determined from the known $H_{\textrm{c2}}(T)$ phase boundary \cite{Cho2009jw}. We note, however, that we are as of yet unable to resolve a phase transition at 10.5 T (corresponding to a $T^{*} = 0.24 T_c$) reported in magnetic penetration depth measurements \cite{Cho2009jw} and possibly also in NMR \cite{Koutroulakis2016jpa}.

\ack
A portion of this work was performed at the National High Magnetic Field Laboratory (NHMFL), which is supported by National Science Foundation Cooperative Agreement No. DMR-11157490 and the State of Florida.

\bibliography{LT28_Fortune_bibliography}

\end{document}